# Trusted CI Experiences in Cybersecurity and Service to Open Science

Andrew Adams
Pittsburgh Supercomputing Center
akadams@psc.edu

Kay Avila
NCSA
kayavila@illinois.edu

Jim Basney
NCSA
jbasney@illinois.edu

Dana Brunson
Internet2
dbrunson@internet2.edu

Robert Cowles
BrightLite Information Security
bob.cowles@gmail.com

Jeannette Dopheide
NCSA
jdopheid@illinois.edu

Terry Fleury
NCSA
tfleury@illinois.edu

Elisa Heymann
University of Wisconsin-Madison
elisa@cs.wisc.edu

Florence Hudson
Independent Consultant
florence.distefano.hudson@gmail.com

Craig Jackson
Indiana University
scjackso@indiana.edu

Ryan Kiser
Indiana University
rlkiser@iu.edu

Mark Krenz
Indiana University
mkrenz@iu.edu

Jim Marsteller
Pittsburgh Supercomputing Center
jam@psc.edu

Barton P. Miller
University of Wisconsin-Madison
bart@cs.wisc.edu

Sean Peisert
Berkeley Lab
sppeisert@lbl.gov

Scott Russell
Indiana University
scolruss@indiana.edu

Susan Sons
Indiana University
sesons@iu.edu

Von Welch
Indiana University
vwelch@iu.edu

John Zage
NCSA
jzage@illinois.edu

## ABSTRACT

This article describes experiences and lessons learned from the Trusted CI project, funded by the US National Science Foundation (NSF) to serve the community as the NSF Cybersecurity Center of Excellence (CCoE). Trusted CI is an effort to address cybersecurity for the open science community through a single organization that provides leadership, training, consulting, and knowledge to that community. The article describes the experiences and lessons learned of Trusted CI regarding both cybersecurity for open science and managing the process of providing centralized services to a broad and diverse community.

# INTRODUCTION

Science projects manage many risks to their mission of reproducible, trustworthy, and productive science. One source of mission risk is typically managed by cybersecurity, e.g., the risk of malicious entities attacking IT infrastructure to further their own ends at the expense of legitimate users or explicitly harming those users. The open science community, with its relatively low need for confidentiality, has traditionally been less concerned with cybersecurity than other communities, such as financial services or medical research. But increasing trends in hacktivism, the politicization of science, and the ability of attackers to monetize data of any sort (via "Ransomware") mean that open science projects are not immune to attack.

This article describes the experiences of Trusted CI, a project initially funded by the National Science Foundation in 2012 and later designated as the NSF Cybersecurity Center of Excellence (CCoE) in 2015. In 2019, Trusted CI is in its seventh year with an objective to address cybersecurity for the United States National Science Foundation (NSF) community, a subset of the global open science community, through a single (virtual) organization that provides training, consulting, and knowledge to that community. Trusted CI also represents a model for providing a specialized skill, cybersecurity, through a centralized organization focused on that skill to serve the whole community. Similar organizations include the UK Software Sustainability Institute [1], the NSF Software Institutes focused on Science Gateways [2] and Molecular Sciences [3], the Department of Energy's Energy Science Network [4], and the NSF Pilot Study for Cyberinfrastructure Center of Excellence [5]. Over the past seven years, Trusted CI has impacted over 250 NSF projects across all seven NSF science directorates [6], helping to secure our nation's scientific cyberinfrastructure (CI) while tailoring cybersecurity to best support scientific missions.

The article describes experiences both in terms of what Trusted CI has learned about cybersecurity for open science and in managing the process of providing centralized services to a broad and diverse community. The former experiences are broadly applicable to science projects and the latter experiences are relevant to other organizations following a similar model of providing focused expertise to the open science community.

The article is organized as follows. We begin with the history of Trusted CI, accompanied by a description of the services it provides. We follow with this article's main contributions: our lessons learned. We continue with a summary of our future vision for the Center. We conclude with related work and acknowledgments.

# HISTORY OF TRUSTED CI

In this Section we provide a brief history of Trusted CI and a summary of its interactions with the US open science community. Subsequent sections provide an analysis of Trusted CI's experiences.

The genesis of Trusted CI is a series of two Scientific Software Security Innovation Institute (S3I2) workshops. The S3I2 workshops, held in 2010 [7] and 2011 [8], included representatives of 35 major NSF-funded projects. The original goal of the workshops was to explore a software institute focused on cybersecurity for the NSF community. One finding from the workshops was that the NSF community faces strong challenges in obtaining access to cybersecurity expertise. Projects are forced to divert their resources to develop that expertise, with the result that they address risks haphazardly, unknowingly reinvent basic cybersecurity solutions, and struggle with interoperability. The workshops further determined that the need for access to expertise was more critical than any new software product.

Based on the finding of the S3I2 workshops, the workshop organizers submitted an unsolicited proposal to NSF to create the Center for Trustworthy Scientific Cyberinfrastructure (CTSC). This proposal was funded in 2012, initiating the project that would eventually become Trusted CI and which we will refer to in this paper as Trusted CI for simplicity. The main role of Trusted CI was to provide cybersecurity expertise to the NSF community. The form of this expertise is described in the following section, but includes engaging with 41 NSF projects, including 9 Large Facilities [9], to aid with cybersecurity challenges; providing cybersecurity training to nearly 300 NSF CI professionals from over 60 projects; and organizing and leading an annual summit to build community and share knowledge in tackling NSF cybersecurity challenges.

In 2015, NSF released a solicitation for a new program, Cybersecurity Innovation for Cyberinfrastructure (CICI), which included a call for an NSF Cybersecurity Center of Excellence (CCoE). The Trusted CI team proposed continued funding for Trusted CI with the NSF CCoE designation and were awarded the designation and three more years of funding (2016-18) from NSF.

Since 2016, Trusted CI, with the NSF CCoE designation, has continued serving the NSF community by convening a working group of leaders from the open science community to develop an Open Science Cyber Risk Profile (in collaboration with ESnet); initiating a Situational Awareness service for the NSF community to inform them of security vulnerabilities and the specific impact on NSF cyberinfrastructure; hosting the annual 100+ attendee NSF Cybersecurity Summit for Large Facilities and Cyberinfrastructure; launching a monthly webinar series covering cybersecurity topics of interest to the NSF community with an average of 30 attendees per event and two thousand subsequent viewings of recordings; providing highly-rated training sessions on cybersecurity topics including identity management, log analysis, secure coding, and related topics at the NSF Cybersecurity Summit, XSEDE, Supercomputing, Indiana University, the Internet2 Technology Exchange, and the eResearch Australasia Conference; and partnering with the newly launched Science Gateways Community Institute (SGCI) to fund half of a security analyst focused on science gateway security. Additional partnerships established to ensure a coherent CI ecosystem included ESnet, Science Gateway Community Institute (SGCI), the Bro Center of Expertise, the Research Security Operations Center, Internet2, InCommon, the National Science Foundation Large Facilities Office, the NSF Pilot Study for Cyberinfrastructure Center of Excellence, the Engagement and Performance Operations Center (EPOC), GÉANT Authentication and Authorisation for Research and Collaboration, the WISE community, and various federal agencies and commercial sector organizations.

In 2019, Trusted CI continued operation via a one-year extension, and the Trusted CI team submitted a five-year proposal in response to a NSF CICI solicitation to continue as the NSF CCoE through 2024. Our five-year vision is to realize an "NSF cybersecurity ecosystem, formed of people, practical knowledge, processes, and cyberinfrastructure, that enables the NSF community to both manage cybersecurity risks and produce trustworthy science in support of NSF's vision of a nation that is the global leader in research and innovation" [10]. We provide more details about our vision and strategic plan in the "Vision for the Future" section.

## TRUSTED CI MODEL FOR EXPERTISE

Compared to the status quo before Trusted CI, in which each NSF project was responsible for obtaining and developing their own cybersecurity expertise, we find that Trusted CI is effective at providing value to the NSF open science community due to the following factors.

**A limited cybersecurity workforce**. The workforce for cybersecurity in general is stretched thin [11], and science projects are challenged to find cybersecurity talent, particularly talent that is also familiar with scientific computing. A Department of Energy Advanced Scientific Computing Advisory Committee Workforce Subcommittee has documented the challenges in finding qualified workforce across a range of computing skills, including cybersecurity [12].

**Limited ability of smaller projects to hire specialized personnel**. A study of cybersecurity budgets [13] at Department of Energy labs found that while there is some variance, their cybersecurity budgets are around 0.5% of their total budget and about 10% of their IT budget. Smaller projects will have a tension if their number of personnel is so small that these percentages no longer translate into full-time or meaningfully fractional part-time employees. The same study found percentages increased as project size decreased, perhaps in response to this tension, but typically small projects need to choose between a specialized cybersecurity position and a more general IT professional.

**Sharing Experiences**. Even when individual projects can find and retain cybersecurity talent, each would only be tackling its slice of the science cybersecurity challenge. The complicated open science ecosystem brings significant challenges in cross-project collaborations and knowledge dissemination. Hard lessons learned by a project are shared haphazardly between projects, if at all. Additionally, important institutional knowledge is often lost when a project is completed, or when key personnel leave the community.

Taken together, these factors combine to lead each open science project to tackle cybersecurity independently, resulting in duplicated mistakes, multiple implementations for security services (e.g., authentication systems) that do not interoperate, and negative impact on the productivity of science by confounding the goals of scientific collaboration, data stewardship, and dissemination of research results.

## EXPERIENCES AND LESSONS LEARNED

We now discuss the experiences of Trusted CI along with the lessons learned related to those experiences, both with regards to cybersecurity for open science, and with regards to operating a center of expertise for open science.

### Engagements

The NSF CICI solicitation in 2015 [14], which funded Trusted CI, called for a Cybersecurity Center of Excellence (CCoE) to conduct security audits and design reviews. We have been conducting such activities for the past seven years as Trusted CI "engagements." Engagements are collaborations between Trusted CI and another project, focused on addressing a cybersecurity challenge of that project. Possible engagement foci include audits and reviews, but can also include: developing a risk-based cybersecurity plan (DKIST, GenApp, LSST, NEO, TransPAC), reviewing existing plans (Array of Things, CC-NIE projects, DataONE, Design Safe, Environmental Data Initiative, Gemini Observatory, HUBzero, IceCube, LNO, NRAO, USAP), making recommendations on software security features (Pegasus, SAGE2, SciGaP, Wildbook), and reviewing software at the code (perfSONAR, Singularity) or architectural level (Globus, HTCondor-CE). Descriptions of our engagements, including impact statements, can be found in our annual reports [15].

### Lessons learned operating a Center of Expertise

**Importance of engagement planning**. Before undertaking the technical work involved in a collaboration, Trusted CI develops an engagement plan in consultation with the engaged project. This document has proven invaluable in ensuring the scope, timeline, committed resources, and outcomes are well understood by both parties. We have found conducting engagements is non-trivial, requiring acumen beyond cybersecurity.

**Engagements require flexibility and innovation**. The exact format of engagements needs to vary depending on the project's culture, goals, challenge, and lifecycle stage. We have learned to innovate and be flexible regarding engagements, experimenting to find ways of making them more efficient. Examples of this innovation include: the introduction of a brief "cybercheckup" at the start of an engagement to evaluate a project's existing cybersecurity program and identify which aspects would benefit most from attention; experimenting with a direct peer-to-peer review between projects that ultimately allows for scale beyond what is possible for Trusted CI to do directly; and ongoing low-effort consulting for newly started projects making a series of design decisions for which they need quick feedback with regards to cybersecurity.

**Tension of sharing**. Trusted CI aims for as broad an impact as possible by sharing the work products of its engagements with the whole NSF CI community. However, projects are sometimes reluctant to share cybersecurity-related information about their project. In these cases, Trusted CI publishes a short summary report of the engagement, while the engaged project receives a complete report containing potentially sensitive details. We have also had some success with the paradigm of developing a project-neutral template to address a relevant cybersecurity issue and then using that to complete the engagement objectives with a project. A template, while not a complete replacement for an actual cybersecurity plan, does serve as a valuable, easily shared resource.

**Be prepared for delays.** Even when a project and engagement approach are well understood, unexpected events (e.g., events that require the engaged project to re-prioritize temporarily) require flexibility in managing the engagement. To adapt to unexpected events, we recognize that our engagement teams will sometimes have spare effort due to being blocked, as well as the occasional need for additional effort. To allow for flexibility, Trusted CI maintains an ongoing task to develop training materials, best practices, and other deliverables with flexible deadlines. This allows staff to be applied to or from those deliverables and time-sensitive engagement tasks.

**Repeated providing of services allows for refinement**. A center of expertise interacts directly with dozens of projects and undertakes similar activities multiple times, something staff embedded in projects would not be able to do without changing positions. This allows Trusted CI staff to experiment and refine those processes. We find engagements are of critical importance as they give us direct and in-depth experience with the challenges facing NSF projects. However, they are labor intensive, consuming much of Trusted CI's resources. Hence, we will continue to mature our engagement methodologies to allow us to perform engagements more efficiently. For example, Trusted CI's Guide to Developing Cybersecurity Programs for NSF Science and Engineering Projects [16] includes over 16 related templates, tools, and resources, and supports NSF CI projects in efficiently building a cybersecurity program to comply with the NSF Cooperative Terms and Conditions for Major Research Equipment and Facilities Construction (MREFC). We developed the Guide as part of our engagement with DKIST, and it was subsequently demonstrated by LSST to be an effective tool for developing a cybersecurity program.

**Engagement impact metrics**. We continue to wrestle with appropriate impact metrics for the engagements. We ask engaged projects to complete a questionnaire at the end of each engagement rating the impact on their project's cybersecurity, including how Trusted CI's assistance compares with other cybersecurity services they may have used. We also follow-up periodically with engaged projects to evaluate impact over time, i.e., after they have implemented Trusted CI recommendations.

**Engagement red flags in terms of lasting impact**. Trusted CI has begun to discern factors that may prevent an engagement from having a deep, sustained impact. Some such factors include the following.

- Project management fails to agree and commit to an engagement plan. We have found a lack of strong commitment by management to be enough of a problem that we will no longer begin an engagement without a mutually agreed-to plan.
- The project has strongly competing priorities. While Trusted CI recognizes projects typically have other tasks they are focused on during an engagement, there are times during a project lifecycle during which the potential for distraction, and hence a lack of participation in an engagement, is increased (e.g., immediately before an initial release). Trusted CI tries to discern this risk and manage it appropriately, e.g., by focusing tightly or delaying an engagement.
- The project lacks application of basic cybersecurity hygiene. While it is true that scientific projects have unusual risks that basic cybersecurity practices do not address well, Trusted CI has determined that applying such basic hygiene, e.g., the SANS Critical Security Controls [17], is a good foundation for nearly any project that has some commodity IT infrastructure. Trusted CI considers if a project applying for an engagement has applied a base hygiene program and either suggests it as a predicate to the engagement or the goal of an initially tightly-focused engagement.

## Lessons regarding Cybersecurity for Open Science

**Management support is critical.** A lack of dedicated resources and/or cybersecurity budget was a problem seen by Trusted CI on numerous occasions and often meant a project was unable to implement any of Trusted CI's recommendations. While Trusted CI does not have a firm minimum metric for resources, we began asking about cybersecurity budgets on engagement applications.

**Standard security controls are still important**. Open science projects rely on a fair amount of standard IT. These projects use email, web browsers, web servers, commodity file sharing, etc., to support their science mission. Common cybersecurity practices (e.g., Securing Commodity IT in Scientific CI Projects: Baseline Controls and Best Practices [18]) work for such IT, and projects should not spend effort re-inventing the wheel.

**Risk-based cybersecurity is needed for specialized scientific IT**. Some IT infrastructure used in support of open science is very uncommon and even unique. Defined practices for cybersecurity do not exist for such IT and development of practices through risk management is necessary. Risk management is time-consuming and takes expertise. Projects are well served to apply standard practices to the greatest extent they can and then prioritize their application of risk management. Trusted CI's Open Science Cyber Risk Profile (OSCRP) [19] provides tools for assessing risks for scientific assets.

## COMMUNITY BUILDING

One of our goals has been to build a community of individuals in NSF projects and supporting organizations who practice cybersecurity to share their experiences, to support each other, and to sustain our work in the event our funding were to end.

We initially attempted to develop an online community using a commercially-hosted web-based service. However, adoption and use by the community lagged and we decided the cost of the service was not merited. We migrated the community over to email lists, which have seen at least as much usage without cost to Trusted CI.

In 2013, Trusted CI re-launched the annual NSF cybersecurity summits after a five-year hiatus. We have continued to organize successful summits for the CI community on a yearly basis, including the introduction of a highly successful Call for Participation process in 2014 to facilitate greater community involvement with the event.

The annual cybersecurity summits provide the community with a valuable opportunity to share best practices, attend practical training sessions, and collaborate on solving common challenges regarding securing NSF-funded facilities and projects. Community evaluations have been overwhelmingly positive.

The summits have become an increasingly important event for Trusted CI development as well. The interaction with the community helps Trusted CI to make new relationships and cultivate opportunities for collaboration with large facilities and projects. Additionally, the knowledge gained from past engagements is communicated at the annual summit, amplifying the impact of the work to the larger community.

In 2016, we started a series of monthly webinars. Initially the content was provided by Trusted CI staff, but it has since expanded to include members of the community. Attendance has grown and has averaged 40 attendees viewing the event live each month, with another 60 viewings of the archives.

## Lessons learned operating a Center of Expertise

**Email lists worked at least as well as web-based services for building community**. Early in the project, we attempted to use a web-based platform to foster community with little success and subsequently shifted to email lists. We speculate that our community is accustomed to using email and in retrospect we should have started with email lists.

**Webinars are an effective way to share expertise**. The webinars were successful both in terms of attendance and in terms of members of the community being willing to present their material. Archival of webinars has built a library of over 30 videos on cybersecurity-related issues.

**The summit has been useful in building a client base**. By fostering interaction with dozens of projects each year, the summit allows Trusted CI to disseminate information about itself and its successes, as well as interact with the community to better understand requirements. Initial contact with many clients has been made through the summit since its inception.

**Summit attendance by NSF staff provides valuable interactions**. Until 2019, Trusted CI held its Summit proximate to NSF (walking distance) to encourage presentation and participation by NSF program officers from a variety of NSF directorates. Like the community interactions, interactions with program officers allow for dissemination of knowledge about Trusted CI and a chance for Trusted CI to understand community requirements. It also gives other Summit attendees an opportunity to interact with their program officers and discuss cybersecurity topics with them face-to-face. However, we have seen declining Summit participation from NSF staff in recent years and increasing costs in the DC area, so we will be experimenting with other Summit locations starting in 2019. Participation in NSF Principal Investigator (PI) meetings and other NSF-organized meetings (such as the NSF Large Facilities Workshop) has provided additional opportunities for NSF staff interactions.

**Venues for delivering training are scarce**. There are few venues that offer opportunities either to provide or receive cybersecurity training targeted at the needs of our community. Many venues face a challenge in making time for specialized topics such as cybersecurity. While Trusted CI has had some success with the PEARC, SC, and XSEDE conferences (primarily with training on secure coding), the Summit remains the main venue for Trusted CI to deliver training. The training at the Summit and other venues has been well received, e.g., 36 of 38 respondents to the 2018 Summit survey answered "yes" when asked if they would participate in future summit training sessions [20]. This leads to the consideration that an event for delivering training to CI professionals by Trusted CI and other projects across a range of specialized topics (e.g., data management, software engineering) could be well received by the community.

## Lessons regarding Cybersecurity for Open Science

**Projects are reticent to discuss cybersecurity incidents**. Often based in embarrassment or reputational concern, projects will initially be unwilling to share experiences regarding their cybersecurity incidents. Over the years, we've seen some erosion of this reticence with sharing of incident experiences [21], but it takes time for the community members to build trust in the community. Education of the community that cybersecurity incidents should not be held against them is also beneficial.

## Other Lessons Learned

**Challenge of commodity IT and underlying organizations**. Every NSF cyberinfrastructure project we have worked with is embedded in and leverages varying degrees of the commodity IT infrastructure, cybersecurity infrastructure, and cybersecurity policies of the university or organization that hosts them. Trusted CI is working on best practices that cover these topics to allow Trusted CI (and other NSF CI projects) to reasonably include them in an assessment and cybersecurity plan without undue effort.

**Cyberinfrastructure has unique security challenges**. In applying best practices from the broader cybersecurity community (e.g., NIST), Trusted CI continues to identify challenges specific to NSF CI, from unique assets such as scientific data and instruments, to challenges such as a close relationship to institutions of higher education and research. CI has a unique threat model, which led to the development of Trusted CI's Open Science Cyber Risk Profile (OSCRP). Additionally, Trusted CI's Guide to Developing Cybersecurity Programs for NSF Science and Engineering Projects provides guidance for addressing challenges unique to NSF CI.

**Strong community ties, operational security expertise, and diverse backgrounds are critical to success**. Since its inception, the Trusted CI team has represented a wealth of operational security experience, strong connections to NSF and other major science projects, and a variety of practical experiences in related domains (e.g., law, risk management) and communities (e.g., software development, scientific, military, corporate, and government).

These differing connections and backgrounds have proven invaluable in being able to initiate and establish the relationships needed to form engagements with diverse scientific communities represented by different NSF projects, as well as bring broader information security best practices to bear.

**Leverage campus resources and expertise when possible**. As we described previously, every NSF CI project with which we have worked is embedded in and leverages varying degrees of the commodity IT infrastructure, cybersecurity infrastructure, and cybersecurity policies of the university or organization that hosts it. Trusted CI has been working to answer questions regarding the degree and circumstances in which projects can leverage existing campus policy and infrastructure. While still not completely understood, some facets of the answers are starting to emerge.

- Commodity services such as vulnerability scanning and licenses for static analysis tools are sufficiently generic to be readily used by projects.
- Campus security offices tend to understand compliance-based security, so a project with HIPAA-covered data or social security numbers will likely find policies or infrastructure they can leverage.
- Due in part to the NSF CC-NIE/IIE program, networks tuned for science (e.g., Science DMZs) are increasingly available and may be of benefit to projects with large data movement needs.
- In general, campuses are not well positioned to provide comprehensive information security plans and programs for complex, large scale, and often multi-institutional science projects.

## RELATED WORK

A crucial goal for Trusted CI is establishing trust and interoperability not only within the NSF community but also with collaborating communities. We seek both to leverage best practices from the broader community as well as to disseminate innovations from Trusted CI and the NSF community. Hence, we have established relationships with communities outside of our key NSF constituency to ensure the success of Trusted CI as a CCoE.

- Department of Energy (DOE): We are well connected with the open science community in DOE through the Energy Science network (ESnet). We collaborate with ESnet on projects with a Science DMZ component and on development of the Open Science Cyber Risk Profile (OSCRP) [19]. In 2019, Sean Peisert joined the Trusted CI team from Berkeley Lab to further strengthen this relationship.
- Science Gateway Community Institute (SGCI): Through SGCI's Incubator program, Trusted CI offers specialized engagements, or consultations, to science gateway developers and operators seeking cybersecurity support. Additionally, Trusted CI presents on relevant cybersecurity topics during SGCI's "bootcamps."
- Bro Center of Expertise: As another large cybersecurity-related project funded by the NSF Office of Advanced Cyberinfrastructure (OAC), collaboration between Trusted CI and the Bro Center was natural. We collaborated on training at the NSF Cybersecurity Summits, engaged with NSF communities with large networking or Bro deployments, and developed documentation and materials, until the Bro Center ended operation in 2018.
- Research Security Operations Center (ResearchSOC): The ResearchSOC, launched in 2018, also under other funding from NSF OAC, provides operational services to the research community. Trusted CI and the ResearchSOC share leadership and coordinate to provide the NSF community with comprehensive cybersecurity leadership and resources.
- Higher Education, Internet2 and InCommon: The NSF Campus Cyberinfrastructure programs demonstrate the continued growth in the portfolio of research support services on campuses, together with the importance of securely connecting campus CI with regional, national, and international CI. Internet2 and InCommon provide a core research network and identity services to campuses and bring the community together to establish standards and share lessons learned. Trusted CI's representation within InCommon leadership (e.g., InCommon Steering Committee and InCommon Technical Advisory Committee) helps to ensure Trusted CI's continued positive impact in this community, building on prior work (e.g., [22]). In 2019, Dana Brunson joined both Internet2 and the Trusted CI team to further strengthen this relationship.
- National Science Foundation (NSF): We have learned there is great value in engaging directly with NSF. We are working closely with the Large Facilities Office to produce cybersecurity guidelines for a future revision of the Major Facilities Guide that is influenced by Trusted CI's cybersecurity planning guide. We also engaged directly with the United States Antarctic Program, which directly operates CI at the Antarctic.
- NSF Pilot Study for Cyberinfrastructure Center of Excellence: This nascent effort is striving to provide a center of expertise regarding CI broadly. Like SGCI, Trusted CI and the Pilot collaborate to offer cybersecurity through the Pilot's activities.
- Engagement and Performance Operations Center (EPOC): The center, established in 2018, is a production platform for operations, applied training, monitoring, and research and education support.
- International Science: Our partnership with the GÉANT Authentication and Authorisation for Research and Collaboration (AARC) project enables EU-US coordination on federated identities for international science. Neil Chue Hong, director of the UK Software Sustainability Institute, serves on our advisory committee, giving us a persistent liaison to science outside of the US. In 2017 and 2018 Trusted CI hosted workshops for the Wise Information Security for collaborating E-infrastructures (WISE) community. WISE is an international community with participants from North America, Europe, Asia, and Australia.
- Other federal agencies and the commercial sector: Through a small number of selected invitations to the annual Cybersecurity Summit, we maintain awareness of other federal agencies and activities in the private sector, as well as allow for the dissemination of our work. Several members of Trusted CI management also serve as PIs on DOE and DHS projects.

## VISION FOR THE FUTURE

The NSF community is large and diverse, encompassing NSF itself, its seven science directorates, over two dozen Large Facilities, and tens of thousands of smaller ephemeral projects. This community is tightly integrated with the higher education institutions and research laboratories that provide administrative homes for projects. The community also collaborates closely with communities from other federal and non-federal agencies, as well as with the international science community.

The diversity of these projects' science missions, combined with the complexities of implementing cybersecurity and open science in tandem, creates a serious cybersecurity challenge. There is no off-the-shelf approach to cybersecurity for open science that the NSF community can adopt. Even Large Facilities, the largest of the NSF projects, struggle to develop tailored approaches.

To address this challenge, an approach is needed to manage risks, while providing both flexibility for project-specific adaptations and access to the necessary knowledge and human resources for implementation. Hence, the Trusted CI vision is for an "NSF cybersecurity ecosystem, formed of people, practical knowledge, processes, and cyberinfrastructure, that enables the NSF community to both manage cybersecurity risks and produce trustworthy science in support of NSF's vision of a nation that is the global leader in research and innovation."

Trusted CI has primary responsibility for bringing the vision for an NSF cybersecurity ecosystem to fruition. Hence, the "mission of Trusted CI is to lead in the development of an NSF cybersecurity ecosystem with the workforce, knowledge, processes, and cyberinfrastructure that enables trustworthy science and NSF's vision of a nation that is a global leader in research and innovation." To accomplish this mission, we organize the future activities of Trusted CI under a set of strategic objectives as follows.

**Build and Disseminate the Needed Knowledge.** Trusted CI will develop and support the adoption of an NSF cybersecurity framework that addresses unique community needs while relating to other broadly known programs such as the NIST Cybersecurity Framework [23] and NIST 800-171 [24]. Trusted CI will be aggressive in evangelizing to projects about the NSF cybersecurity framework, related cybersecurity resources, and workforce development opportunities. Additionally, Trusted CI will continue to organize the annual NSF Cybersecurity Summit, online discussions, and communication forums to continue to mature and grow the community. In these efforts, Trusted CI will continue to espouse a flexible approach to cybersecurity, which balances baseline practices with risk management emphasizing the mission of scientific research.

**Sustain the Community.** Trusted CI will continue to improve its delivery of NSF- and project-funded engagements to most effectively and efficiently meet the needs of the NSF community. It will also continue to explore Engagements funded by projects as a means of achieving the goal of financial sustainability. Trusted CI will lead the definition and tracking of cybersecurity community metrics to: 1) measure the impact of Trusted CI; 2) enable community members to benchmark their efforts in relation to the cybersecurity efforts of other community members; and 3) demonstrate the maturation of the NSF cybersecurity ecosystem over time.

**Secure Cyberinfrastructure.** Trusted CI will improve the security of NSF cyberinfrastructure by developing secure software engineering and secure coding practices. It will also build a national community around cybersecurity for research by coordinating events that build trust across NSF and collaborating organizations both nationally (e.g., Department of Energy, National Institutes of Health) and internationally (e.g., Large Hadron Collider, Square Kilometer Array, Laser Interferometer Gravitational-Wave Observatory).

**Foster the Workforce and Collaborations.** Trusted CI will continue to provide high-quality training for the NSF community and will continue efforts to make students and non-NSF professionals aware of the NSF cybersecurity ecosystem and the opportunities to work in cybersecurity and to enable science, an exciting combination. Trusted CI will also continue work to increase the representation of minorities and underrepresented groups in the NSF cybersecurity ecosystem, whose demographics at NSF Cybersecurity Summits indicate a white male majority, to further bolster the workforce. Trusted CI will continue outreach to higher education information security officers and research facilitators to enable them to help NSF projects with cybersecurity challenges.

As part of realizing this vision, Trusted CI has launched two new activities in 2019. Trusted CI established a fellowship program to further address the scaling challenge of impacting the tens of thousands of NSF funded projects, and Trusted CI established a transition to practice (TTP) program to foster the transition of NSF-funded cybersecurity research and development into practice in the NSF community and convey unmet cybersecurity requirements back to the research and development communities as targets for future research.

## CONCLUSIONS

The experience we have gained while operating Trusted CI since 2012 has helped us develop effective methods of addressing cybersecurity for open science. As the NSF Cybersecurity Center of Excellence, Trusted CI is a model for providing national-scale expertise to the scientific community, working collaboratively with science projects to fill gaps and disseminate knowledge via training, consulting, webinars, summits, and email lists. We hope the experiences and lessons learned presented here are helpful to organizations that are providing support for cybersecurity or other focused topic areas (e.g., software engineering, high performance networking, etc.) to the scientific community.

For additional Trusted CI resources, visit https://trustedci.org/.

## ACKNOWLEDGMENTS

We are grateful to the Trusted CI advisory committee for their advice and direction over the years. Current and former advisory committee members are: Tom Barton, David Halstead, Neil Chue Hong, Don Middleton, Nick Multari, Nancy Wilkins-Diehr, and Melissa Woo. We thank John McGee for his contributions to the original CTSC proposal. We thank former Trusted CI team members Jared Allar, Rakesh Bobba, Randy Butler, Patrick Duda, Randy Heiland, Scott Koranda, Warren Raquel, and Amy Starzynski Coddens. We thank our NSF program officers (Anita Nikolich and Kevin Thompson) for their guidance and support. This material is based upon work supported by the National Science Foundation under grant numbers 1234408 and 1547272.

## REFERENCES

[1] Crouch, Stephen; Chue Hong, Neil; Hettrick, Simon; Jackson, Mike; Pawlik, Aleksandra; Sufi, Shoaib; Carr, Les; De Roure, David; Goble, Carole; Parsons, Mark, "The Software Sustainability Institute: Changing Research Software Attitudes and Practices," *Computing in Science & Engineering*, vol.15, no.6, pp.74,80, Nov-Dec 2013. DOI: 10.1109/MCSE.2013.133.

[2] Science Gateways Community Institute. [Online]. https://sciencegateways.org/

[3] Molecular Sciences Software Institute. [Online]. http://molssi.org/

[4] Energy Science Network (ESnet). [Online]. http://es.net/

[5] Pilot Study for a Cyberinfrastructure Center of Excellence. [Online]. https://www.nsf.gov/awardsearch/showAward?AWD_ID=1842042

[6] Jeannette Dopheide, John Zage, and Jim Basney, "The Trusted CI Broader Impacts Project Report" 2018. [Online]. http://hdl.handle.net/2022/22148

[7] William Barnett, Jim Basney, Randy Butler, and Doug Pearson, "Report of the First NSF Workshop on Scientific Software Security Innovation Institute," October 2010. [Online]. http://security.ncsa.illinois.edu/s3i2/

[8] William Barnett, Jim Basney, Randy Butler, and Doug Pearson, "Report of the Second NSF Workshop on Scientific Software Security Innovation Institute," November 2011. [Online]. http://security.ncsa.illinois.edu/s3i2/

[9] Large Facilities Office (LFO). [Online]. https://www.nsf.gov/bfa/lfo/

[10] V. Welch, J. Basney, C. Jackson, J. Marsteller, and B. Miller, "The Trusted CI Vision for an NSF Cybersecurity Ecosystem and Five-year Strategic Plan (2019-2023)," Trusted CI, Apr. 2018 [Online]. http://hdl.handle.net/2022/22178

[11] M. Suby and F.Dickson. "The 2015 (ISC)2 Global Information Security Workforce Study," 2015 [Online]. Available: https://www.isc2.org

[12] IAdvanced Scientific Computing Advisory Committee. "ASCAC Workforce Subcommittee Letter," 2014 [Online]. Available: https://science.energy.gov/~/media/ascr/ascac/pdf/charges/ASCAC_Workforce_Letter_Report.pdf

[13] S. Russell, B. Cowles, C. Jackson, "Cybersecurity Budgets," NSF Cybersecurity Summit, Arlington, VA, August 2016.

[14] NSF Solicitation 15-549. [Online]. https://www.nsf.gov/publications/pub_summ.jsp?ods_key=nsf15549&org=NSF

[15] Trusted CI Annual Reports. [Online]. https://trustedci.org/reports

[16] "Guide to Developing Cybersecurity Programs for NSF Science and Engineering Projects," v1, Center for Trustworthy Scientific Cyberinfrastructure. August 2014 [Online]. https://trustedci.org/guide

[17] Center for Internet Security, Critical Security Controls. [Online]. https://www.sans.org/critical-security-controls/

[18] "Securing Commodity IT in Scientific CI Projects: Baseline Controls and Best Practices." August 2014 [Online]. https://trustedci.org/guide/docs/commodityIT

[19] Sean Peisert, Von Welch, Andrew Adams, RuthAnne Bevier, Michael Dopheide, Rich LeDuc, Pascal Meunier, Steve Schwab, and Karen Stocks. 2017. "Open Science Cyber Risk Profile (OSCRP)," v1.2. March 2017. [Online]. http://hdl.handle.net/2022/21259

[20] Andrew Adams, Jeannette Dopheide, Mark Krenz, James Marsteller, Von Welch, and John Zage. "Report of the 2018 NSF Cybersecurity Summit for Cyberinfrastructure and Large Facilities." 2018 [Online]. http://hdl.handle.net/2022/22588

[21] Craig Jackson, Amy Starzynski Coddens. "Report of the 2015 NSF Cybersecurity Summit for Cyberinfrastructure and Large Facilities: Large Facility Cybersecurity Challenges and Responses." 2015 [Online]. http://hdl.handle.net/2022/20539

[22] W. Barnett, W, V. Welch, A. Walsh and C.A. Stewart. "A Roadmap for Using NSF Cyberinfrastructure with InCommon," 2011 [Online]. http://hdl.handle.net/2022/13024

[23] "Framework for Improving Critical Infrastructure Cybersecurity," National Institute of Standards and Technology, Feb. 2014 [Online]. Retrieved April 1, 2019 from https://www.nist.gov/cyberframework

[24] A. R. R. (nist), A. K. D. (nist), A. P. V. (nara), A. M. R. (nara), and A. G. G. (ida), "SP 800-171 Rev. 1, Protecting CUI in Nonfederal Systems and Organizations | CSRC." [Online]. Retrieved April 1, 2019 from https://csrc.nist.gov/publications/detail/sp/800-171/rev-1/final